# ProtFract: A server to calculate interior and exterior fractal properties of proteins: Case study with Ras superfamily protein structures

Charudatta Navare, Anirban Banerji*

Bioinformatics Centre, University of Pune, Pune-411007, Maharashtra, India

E-mail: anirbanab@gmail.com

## Abstract

**Motivation:** Protein surface roughness is fractal in nature. Mass, hydrophobicity, polarizability distributions of protein interior are fractal too, as are the distributions of dipole moments, aromatic residues, and many other structural determinants. The open-access server ProtFract, presents a reliable way to obtain numerous fractal-dimension and correlation-dimension based results to quantify the symmetry of self-similarity in distributions of various properties of protein interior and exterior.

**Results:** Fractal dimension based comparative analyses of various biophysical properties of Ras superfamily proteins were conducted. Though the extent of sequence and functional overlapping across Ras superfamily structures is extremely high, results obtained from ProtFract investigation are found to be sensitive to detect differences in the distribution of each of the properties. For example, it was found that the RAN proteins are structurally most stable amongst all Ras superfamily proteins, the ARFs possess maximum extent of unused hydrophobicity in their structures, RAB protein interiors have electrostatically least conducive environment, GEM/REM/RAD proteins possess exceptionally high symmetry in the structural organization of their active chiral centres, neither hydrophobicity nor columbic interactions play significant part in stabilizing the RAS proteins but aromatic interactions do, though cation-π interactions are found to be more dominant in RAN than in RAS proteins. The specific π–π and cation-π interactions are found to be of exceptional significance to stability of Ras superfamily proteins. Due to distinct class-specific nature of π–π and cation-π interaction symmetries, they are found to be the best classifiers to segregate the highly similar Ras superfamily proteins from one another.

**Availability:** ProtFract is freely available online at the URL: http://bioinfo.net.in/protfract/index.html

**Contact Information:** protfract@gmail.com

## 1. Introduction

Several protein properties embody symmetry of self-similarity. Symmetries observed in crystalline structures can be described with translations, rotations, and reflections; symmetry of self-similarity, on the other hand, manifests itself through scale invariance. Self-similar forms are composed of subunits that resemble the overall object. If the Hausdorff dimension (Hausdorff, 1919) of a self-similar object is found to be greater than its topological dimension, the object is called a fractal-object. What makes this information relevant in the field of Bioinformatics is the fact that proteins are fractal-objects, as established by works of last three decades.

Protein surface roughness demonstrates fractal symmetry (Lewis and Rees, 1985). Mass, hydrophobicity and polarizability distribution of protein interior are fractal (Banerji and Ghosh, 2009b). Many other protein properties (main-chain connectivity profile, potential energy profile, dipole moment distribution etc. (Banerji and Ghosh, 2011; and references therein)) – are fractal too. However, a computational resource that calculates fractal properties of protein interior and exterior – is difficult (if not impossible) to find.

ProtFract, a freely accessible online server, calculates the fractal dimension of many of the protein properties that embody the symmetry of self-similarity in their distribution.

## 2. Protfract features

Protfract offers services of two broad types; services that quantify protein interior fractal-properties and services that quantify protein exterior fractal-properties.

### 2.1 Protein interior analyses

Proteins are 'complex mesoscopic systems' (Karplus, 2000). The standard compact-object description of proteins (characterized by small-amplitude vibrations and by low-frequency Debye density of states) cannot account for their 'non-idealistic' behavior (de Leeuw et al., 2009). Hence, rigorous investigations of protein property distributions may require a methodology that treats proteins from the perspective of a non-idealistic (complex) system. The fractal dimension(FD) and correlation dimension(CD) (another type of FD) are two such robust and consistent constructs – as has been established recently (Banerji and Ghosh, 2011).

To calculate the self-similarity in inhomogeneous distributions (Banerji and Ghosh, 2009a) of mass, hydrophobicity and polarizability, ProtFract resorts to FD-based analysis. ProtFract's CD-based analyses, on the other hand, quantify the self-similarity in dependencies (alternatively, correlations) amongst biophysical properties (Banerji and Ghosh, 2011). ProtFract CD-suit quantifies the extent of self-similarity in – dipole moment distributions, (active) chiral centre distributions, aromatic residue distributions, charged residue distributions, and in distribution of interactions between aromatic residues and positively charged residues.

Input for ProtFract interior services is accepted in the form of PDB files of respective proteins. Output of ProtFract interior services is provided as a (zipped) set of comma separated Excel files, text files and figure files.

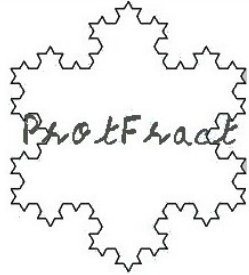

(a)

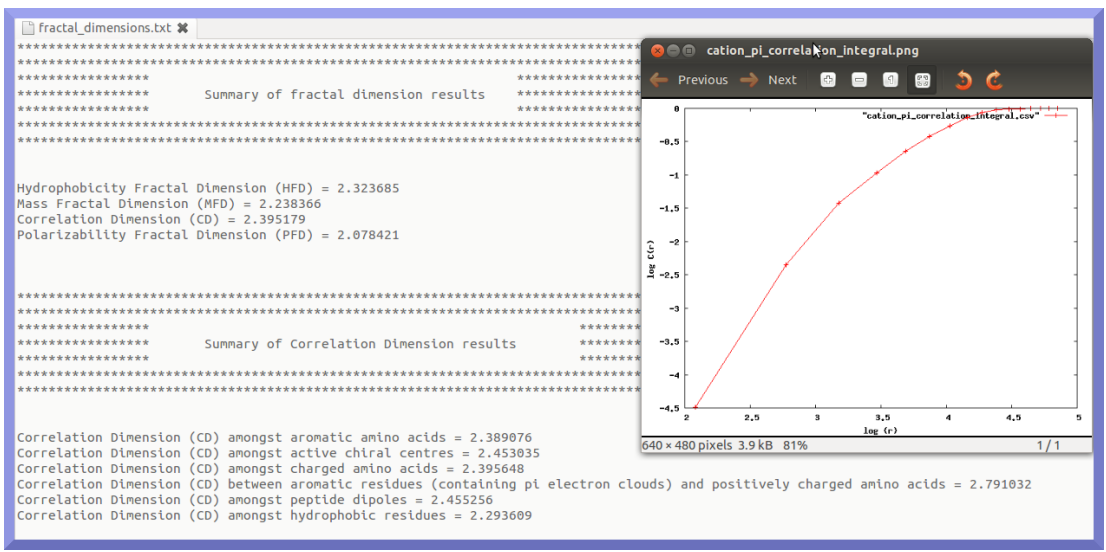

(b)

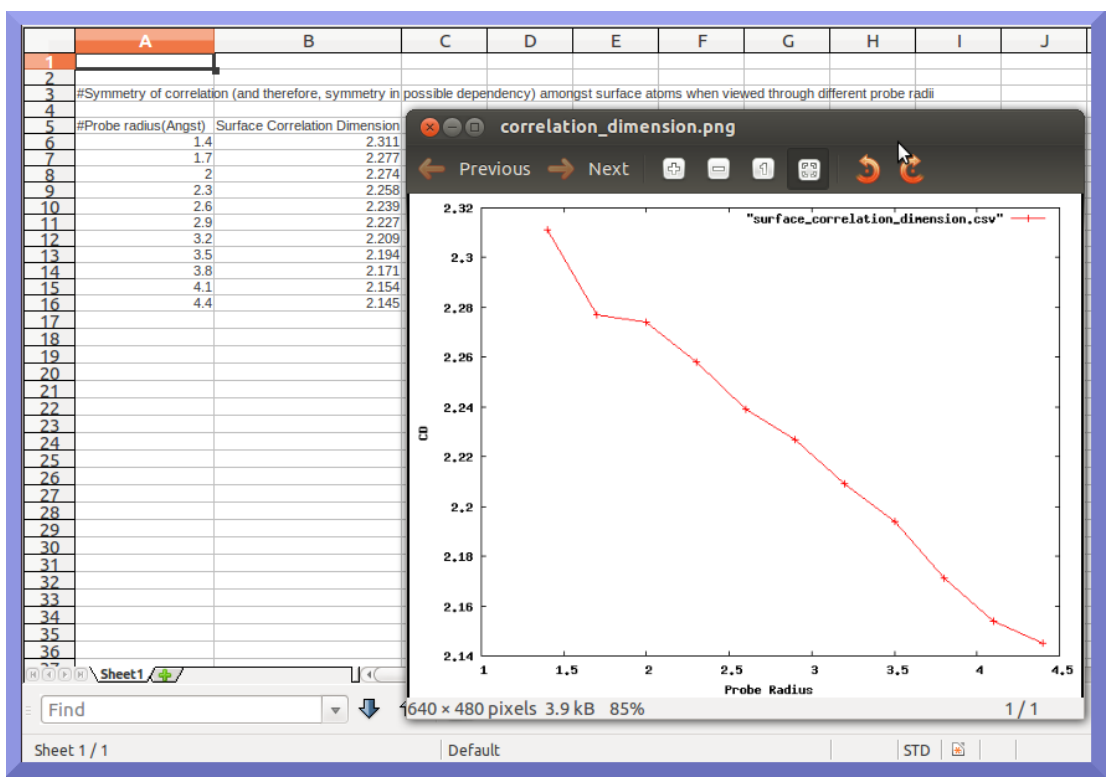

(c)

**Figure 1**

(a) a screen-shot of ProtFract home page. (b) sample outout of protein interior analyses.

(c) sample output of protein exterior analyses

**2.2 Protein exterior analyses**

Roughness of protein surfaces introduces an additional complexity in protein-binding investigations. Past studies have established the utility of surface-FD in the context of studying protein-ligand interactions (Pettit and Bowie, 1999) and protein-protein interactions (Aqvist and Tapia, 1987). Thus, ProtFract provides two facilities to calculate surface-roughness with FD. In the first, surface-FD of entire protein surface is calculated, which presents the user with an idea of the global corrugation profile of the protein. The second surface-FD service presents information about roughness of a particular surface patch (say, the patch surrounding the active site) on protein surface. One may quantify the self-similarity in surface roughness with respect to the size of an approaching probe too. Building upon this idea, a previous study (Choi and Lee, 2000) employed CD to quantify the protein surface roughness. Applying this algorithm ProtFract offers users the information regarding CD amongst surface atoms of the proteins, which may prove beneficial for studying of solvent mobility near protein surface. To obtain the molecular surface information, 'Surface Racer' (Tsodikov et al., 2002) is used.

Input for ProtFract exterior services can be provided in two ways. To obtain information regarding the global texture, one needs to input the entire PDB coordinate file for that protein. To obtain information regarding surface roughness of specific local patch, user needs to input the chain identifier and residue number information. Output of ProtFract exterior services is presented as a (zipped) set of text files and figure files.

### 3. Results: Segregating Ras superfamily proteins with comparative structural analyses.
### 3.1. Background

The Ras superfamily consists of large set of structurally and functionally conserved small monomeric GTP-binding proteins with molecular masses varying within a small range, 20 to 40 kDa (Takai et al., 2001). They act as molecular switches alternating between a GTP-bound ON state (where proteins acquire activation by binding to GTP) and a GDP-bound OFF state (where proteins suffer inactivation by hydrolysis of GTP to GDP) (Vetter and Wittinghofer, 2001; Corbett and Alber, 2001; Wennerberg et al., 2005). Ras superfamily have high sequence identity (40-85%) among them (Vigil et al., 2010), with the bulk of this identity derived from four conserved domains required for the recognition of GDP, GTP and for GTPase activity (Valencia et al., 1991). Not surprisingly, structures of proteins belonging to RAS superfamily are found to have lot of commonalities among themselves, whereby they are found to populate the same fold 'P-loop containing nucleoside triphosphate hydrolases' belonging to the structural class alpha/beta (for all the cases for which structural domain information could be found). Due to such overwhelming commonality, the structure comparison results among RAS superfamily structures obtained from DALI (REF) registered very high (> 20.0) Z-score in every case.

These observations lead us to a paradoxical situation, where one notes high levels of conservation in sequence, structure and function paradigm across Ras superfamily proteins, and yet, one finds that each of the individual proteins have unique functions and preferred targets. This paradox can only be resolved if one hypothesizes that differences of functionalities in Ras superfamily proteins can be equated to subtle differences in biophysical properties in them. Such a hypothesis demands that subtle differences in biophysical properties among Ras superfamily proteins can be quantified at the first place, which is difficult to achieve because very high proximity in structural organization space implies (an equally) close proximity in distribution of biophysical properties too. Thus the classification scheme requires to be sensitive to detect small dissimilarities in biophysical properties, whereby patterns can be deciphered in the closely-related and yet subtly-different distributions. The patterns in differences among biophysical properties among Ras superfamily proteins serve as the important first step to understand how, even with extraordinarily high structural similarities among them the Ras superfamily proteins can function in distinctly characteristic ways.

However, even before attempting to tackle this challenging problem one notes that proteins are not classical solid objects, but are fractal objects. Hence, ProtFract server based analysis was carried out on the Ras superfamily proteins.

### 3.2. Segregation of Ras superfamily proteins with biophysical means

**3.2.1. Materials:** Considering the dataset provided in a recent work (Magis et al., 2010), analysis was conducted on all the non-redundant protein structures available in PDB (Berman et al., 2003). These included proteins from RAB, RAS, ARF, RHO, RAN and GEM/REM/RAD domains. Proteins from RGK and SAR domains could not be considered owing to certain ambiguities in their structural information.

**3.2.2. Method:** Structural information of each protein was provided as input to both interior and exterior analysis of the server ProtFract. The Help section of ProtFract server presents a more detailed account of algorithms employed.

**3.2.3. Results:** While the ProtFract server provides the output of biophysical analyses for a single protein, such an account for protein biophysical properties will not be helpful to characterize the family-specific nature of biophysical properties of Ras superfamily proteins. In order to compare and contrast these, property-wise averaging was undertaken for each of the ProtFract output properties for each of the Ras superfamily members. The comparative results are presented below for sets of related properties.

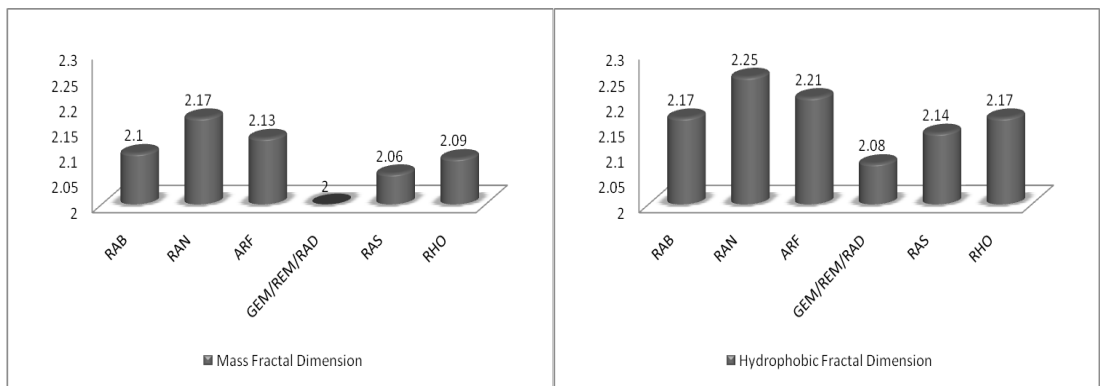

**Figure-2**          **Figure-3**

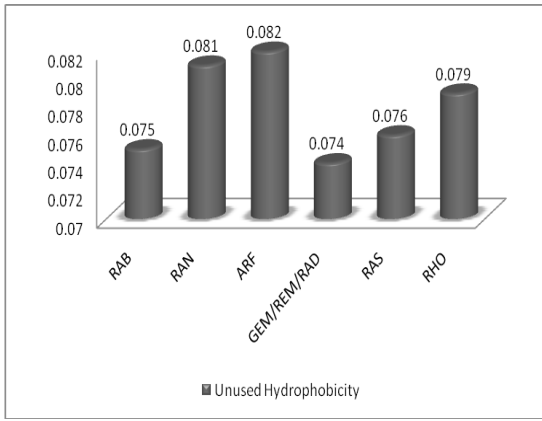 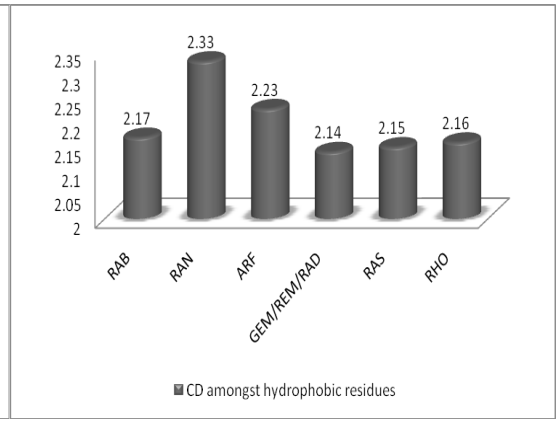

**Figure-4**  **Figure-5**

**Figure Legend for Figure-2 to Figure54 (alongside small discussion):** Comparative analysis of the mass fractal dimension(MFD), hydrophobicity fractal dimension(HFD), unused hydrophobicity(UH(=HFD-MFD)) and correlation dimension(CD) among hydrophobic residues (ALA, VAL, LEU, ILE, MET, PHE, TRP, CYS). The Ras superfamily GTPases are all small proteins with ~20 kDa G domain (Ras residues 5-166). Since all of them are small in size with less mass content, extent of space-filling symmetry of mass across them is also found to be low, whch accounted for all of them registering low and yet, comparable magnitudes of MFD. The RAN family of proteins are found to have more compactness in them, as compared to Ras superfamily proteins; GEM/REM/RAD proteins are found to be least compact of all of them. Hydrophobic effects are found to be more space-filling than mass distribution across all the Ras families. Though hydrophobicity is more space filling in RAN (HFD$_{RAN}$=2.25 and CD$_{RAN\text{-}Hydroph\text{-}Residues}$=2.33) than that in ARF(HFD$_{ARF}$=2.21 and CD$_{ARF\text{-}Hydroph\text{-}Residues}$=2.23), the unused hydrophobicity is found to be more in ARF than that in RAN.

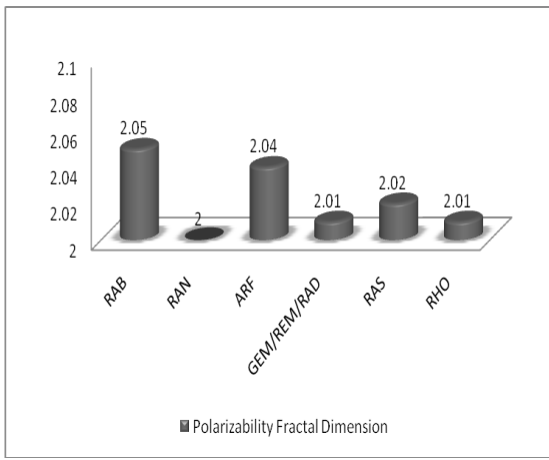 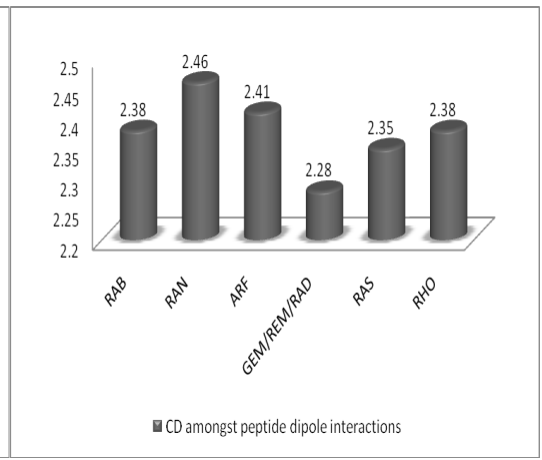

**Figure-6**  **Figure-7**

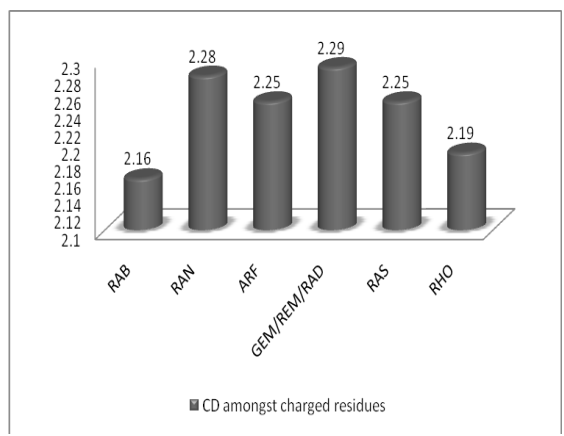

**Figure-8**

**Figure Legend for Figure-6 to Figure-8(alongside small discussion):** Lower residual polarizability implies lower probability of protein interior dielectric constant, which in turn, implies more conducive environment for electrostatic interactions of numerous kinds. Probability of a protein structure acquiring higher stability is therefore dependnt upon its possessing high magnitude of HFD or low magnitude of PFD or an optimal balance between the two. As demonstrated by comparative analysis (**Fig-6**), though the electrostatic environment in all the Ras superfamily proteins are highly conducive (whereby the polarizability fractal dimension(PFD) in them is recorded to be consistently $\leq 2.05$, viz. effect of residual charge separation is extremely restricted in space), RAB proteins are electrostatically least conducive while RAN proteins are found to have best electrostatic milieu among all the Ras superfamily proteins. It is therefore unsurprising to note that extent of symmetry of self-similarity in space-filling nature of (backbone) dipole-dipole interactions is distinctly higher in RAN proteins than that in any others (**Fig-7**). Just as the symmetry of spatial correlation among peptide-dipoles could be measured by calculating its CD, symmetry of spatial correlation in residual coulombic interaction could be measured by quantifying the global symmetry in spatial proximity ($C^\beta$-$C^\beta \leq 8$Å) of (ASP, GLU) in one hand and (LYS, HIS, ARG) in the other (whereby, there are 6(=3+3) attractive interactions and 4(=3(=2+1) + 1) repulsive interactions). Though residual electrostatics driven interactions are found to be predictably high within the RAN proteins, GEM/REM/RAD proteins are found to be maximally dependent on residual electrostatics (**Fig-8**). The RHO proteins, though possessing identical dielectric profile as that of GEM/REM/RAD proteins (for both PFD=2.01), could be observed to have better symmetry among their peptide dipoles than that in the later, but residual electrostatics in RHO is found to be proportionately lower than that in GEM/REM/RAD proteins.

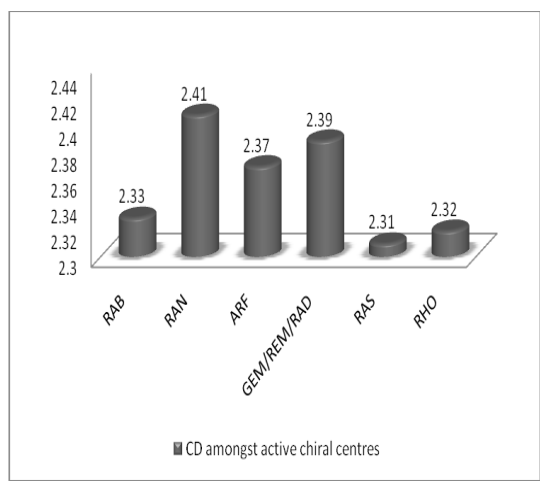

**Figure-9**

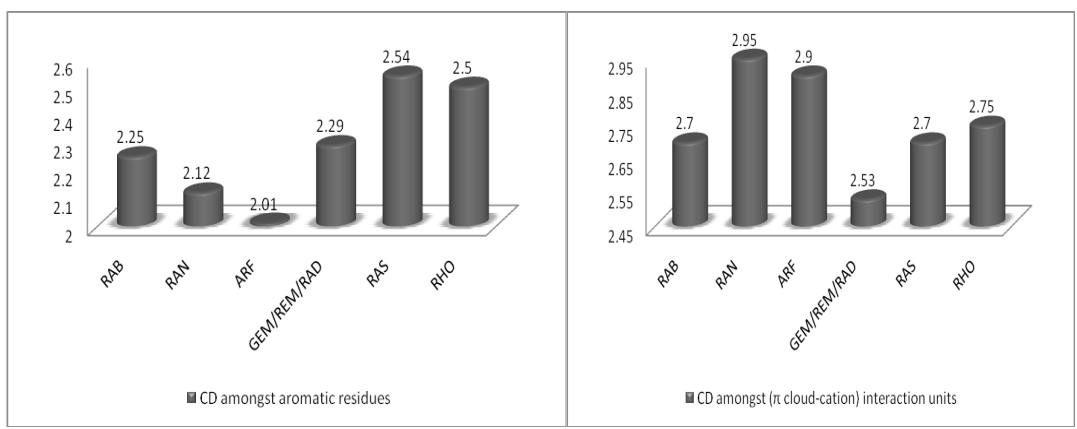

**Figure-10**  **Figure-11**

**Figure Legend for Figure-9 to Figure-11(alongside small discussion):** Though the structures of Ras superfamily proteins are highly similar, differences could be detected in the manner how active chiral centres (all the $C^\alpha$s alongside $C^\beta$s of ILE and THR) are accommodated in them. While the structural organization of RAN proteins is recorded to have maximum symmetry in spatial arrangement of active chiral centres (**Fig-9**), most unexpectedly, the GEM/REM/RAD proteins though having the least mass distribution among all Ras superfamily proteins, are found to possess the second-largest symmetry in their spatial organization of active chiral centres. Ras proteins are detected to have minimum symmetry in the way how active chiral centres are positioned in them.

Among all the properties widest variations could be detected in the symmetry of aromatic-aromatic side-chain interactions (**Fig-10**). While the π cloud-π cloud interaction symmetry is found to be maximum in RAS proteins, ARF proteins recorded least such symmetry in them. The cation-π interaction symmetry (**Fig-11**), surprisingly, registered a different trend for itself, whereby not RAS but the RAN proteins recorded the maximum symmetry, with ARF proteins (which possess least π-π symmetry in them) registering the second-highest space-filling nature. The reasons and implications of these are provided in details in 'Discussion' section.

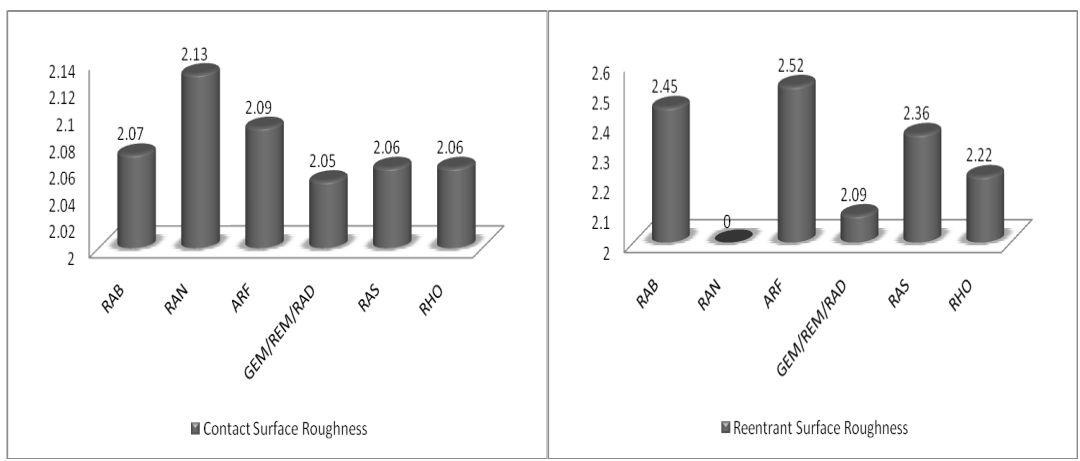

**Figure-12**    **Figure-13**

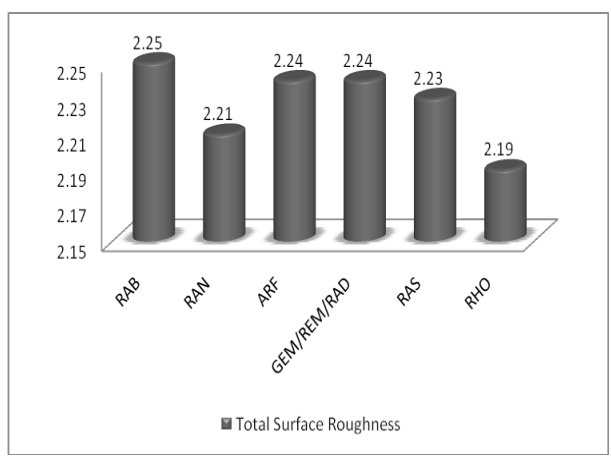

**Figure-14**

**Figure Legend for Figure-12 to Figure-14(alongside small discussion):** Protein surface roughness is an important descriptor of molecular surface, it is the third-best property (after shape complementarity and charge complementarity) to influence binding strategies of a protein (Pettit et al., 2007). Thus symmetry of roughness of contact, reentrant and total surface (Connolly, 1983) were calculated to investigate the exterior characteristics in details.

Due to their small sizes the Ras superfamily proteins did not register high magnitude of contact surface roughness and all of them (barring RAN proteins) recorded similar contact surface roughness (**Fig-12**). Statistically significant number of data points could not be found for most Ras superfamily reentrant surfaces (RAN surfaces in particular). The large variation of obtained results (**Fig-13**), therefore, may not be reflective of the true biological reality. Quantification of the total surface roughness, however, showed unambiguously that surface roughness of Ras superfamily proteins are very similar (**Fig-14**). Such similarity implies that *difference* in recognition strategies of Ras superfamily proteins, if any, will certainly not be rooted in their surface roughness but in other (shape or charge-related) descriptors.

**Discussion:**

The broad patterns observed in the obtained results can be used to infer certain general facts about structural organization of Ras superfamily proteins.

1) **(Possible) superior stability profile of RAN proteins:** The fact that RAN family of proteins registered highest self-similarity in space-filling symmetry in hydrophobicity distribution (**Fig-3**, **Fig-5**) and lowest self-similarity in space-filling symmetry in polarizability distribution (**Fig-6**) suggest unambiguously that RAN family of proteins are most structurally stable among all the Ras superfamily proteins.

2) **Coulombic interaction-driven structural organization of GEM/REM/RAD:** Less polarizability implies less dielectric constant, which in turn, implies a more conducive environment for electrostatic interactions. The fact that both $HFD_{GEM/REM/RAD}$ (**Fig-3**) and CD among hydrophobic residues of GEM/REM/RAD (**Fig-5**) are lowest among Ras superfamily proteins, establishes that fact that GEM/REM/RAD proteins do not bank on hydrophobicity to ensure their stability. However, for them the low PFD (**Fig-6**) accounts for highest CD among charged residues (**Fig-8**). Then again, the symmetry in (peptide)dipole-dipole interactions in them (**Fig-7**) is found to be least space-filling in comparison to that observed in other Ras superfamily proteins. Thus one may attribute the stability of GEM/REM/RAD proteins (entirely) to superior profile of coulombic interactions in them.

3) **Distinctly different organizational schemes of RAS proteins:** RAS proteins are found to possess atypical characteristics in their structural organization. In one hand, consensus patterns obtained from **Fig-3** and **Fig-5** suggest unambiguously that hydrophobicity distribution symmetry in RAS proteins is lower than that found in RAN, ARF, RAB and RHO proteins, which implies that stability of RAS proteins cannot be attributed to hydrophobicity distribution therein. On the other hand, though the environment for electrostatic interactions in RAS is found to be (slightly) more conducive than that in RAB and in ARF (**Fig-6**), the symmetry of dipole-dipole interactions in RAS proteins are found to be weaker than that in RAN, ARF, RAB and RHO proteins (**Fig-7**). Similarly, though the interaction symmetry amongst charged residues in RAS matched that in ARF, such symmetry in Coulombic interactions is found to be not as strong as one observed in GEM/REM/RAD and RAN proteins (**Fig-8**).

However RAS proteins are found to derive much of their structural stability from the symmetry of π-π interactions in them, which is found to be the highest amongst all the Ras superfamily proteins. Interaction between two aromatic side-chains typically contributes -0.6 to -1.3kcal/mol to protein stability (Serrano et al., 1991). It is known (Burley and Petsko,

1985) that specific and energetically favorable π-π interactions in peptides may substantially help the peptides in stabilizing and in maintaining their functionally active shape. Likewise, in proteins, aromatic interactions (and aromatic clusters especially) have been reported to have significant roles in ensuring thermal stability (Georis et al., 2000; Kannan and Vishveshwara, 2000). From the set of obtained results it seems logical to hypothesize that RAS proteins are more dependent on aromatic interactions (than on hydrophobicity or on Coulombic interactions) for their stability. Space-filling symmetry of cation-π interactions in RAS proteins, however, is found to be weaker than that in RAN, ARF and RHO proteins, implying clearly that it is the spatial proximity and specific geometry of aromatic side-chains (and not the interactions of these aromatic groups with cationic side-chains) that hold unique importance for the RAS proteins. We found (only) two previous works (Neuwald et al., 2003; Thomas et al., 2007) that commented upon the importance of aromatic interactions in Ras superfamily proteins, but in both cases these assessments were based upon particular observations under particular contexts; thus, they could not be helpful to ascertain the general trends (and preferences) that compare and contrast Ras superfamily classes.

4) **Exceptional significance of cation-pi interactions in Ras superfamily proteins:** Just like the case of aromatic interactions, wide variations could be recorded in extent of symmetries for cation-π interactions in Ras superfamily proteins. The stabilization energy of cation-π interactions originates from electrostatic interaction between the cation and high-electron density π-orbitals from the aromatic moiety of aromatic residues (Burghardt et al., 2002; Petersen et al., 2005). The distinctive aspects of observed cation-π symmetries in Ras superfamily of proteins are that; *first*, they are typically more space-filling than symmetry found in the distribution of any of the other properties - which strongly suggests the possible roles of cation-π interactions in stabilizing the Ras superfamily proteins; and *second*, cation-π interaction symmetries registered little dependence on the degree of presence of either the π-π interactions or the Coulombic interactions, and yet, they demonstrated strong symmetries that varied distinctly from one class to another in proteins belonging to Ras superfamily. To exemplify, though (**Fig-10**) the symmetry existing in π-π interactions in ARF proteins is recorded to be minimum ($CD^{\pi\text{-}\pi}_{ARF}$ = 2.01, viz. almost non-existent in comparison to $CD^{\pi\text{-}\pi}_{RAS}$ = 2.54 and $CD^{\pi\text{-}\pi}_{RHO}$ = 2.50) and though (**Fig-8**) the symmetry existing in interaction profile amongst charged residues in ARF proteins is recorded to be the same as that in RAS proteins ($CD^{charged\text{-}residues}_{ARF}$ = $CD^{charged\text{-}residues}_{RAS}$ = 2.25), the cation-π symmetry in ARF proteins recorded an astoundingly high symmetry ($CD^{cation\text{-}\pi}_{ARF}$ = 2.90). Similar observations can be made for RAN and RAB proteins too, whereas RHO proteins recorded similarity with RAS in the comparative symmetry analyses of $CD^{\pi\text{-}\pi}$, $CD^{charged\ residues}$ and $CD^{cation\text{-}\pi}$ interactions.

## 4. Conclusion

Since an ever-increasing number of studies are viewing proteins as 'complex-systems', construction of an accurate (already validated) and freely available web-server to calculate the fractal-properties of proteins assumes pivotal importance. Using FD and CD, the server ProtFract can help users in quantifying symmetries in organization of various biophysical properties that describe protein interior and exterior. Though essentially statistical in nature, both FD and CD are found to be sensitive to detect and quantify the finer differences that exist in organizational schemes of biophysical properties in highly similar proteins across Ras superfamily domains. With such a quantification of symmetries, the local level (residual) interactions and global level (quaternary) structural stability aspects – could be addressed with a uniform scale-invariant symmetry measure.

**Acknowledgement:**

Anirban Banerji thanks COE-DBT scheme, Govt. of India, for supporting him.